\documentclass[aps,prl,reprint,superscriptaddress,twocolumn]{revtex4-1}

\usepackage{xcolor}
\usepackage{graphicx}
\usepackage[T1]{fontenc}
\usepackage{lineno}

\newcommand{\STEREO}{STEREO}
\begin{document}

%\preprint{}

\title{Sterile Neutrino Constraints from the \STEREO{} Experiment with 66 Days of Reactor-On Data}



\author{H.~Almaz\'{a}n}
\affiliation{Max-Planck-Institut f\"ur Kernphysik, Saupfercheckweg 1, 69117 Heidelberg, Germany }
\author{P.~del Amo Sanchez}
\affiliation{Univ. Grenoble Alpes, Universit{\'e} Savoie Mont Blanc, CNRS/IN2P3, LAPP, 74000 Annecy, France}
\author{L.~Bernard}
\affiliation{Univ. Grenoble Alpes, CNRS, Grenoble INP, LPSC-IN2P3, 38000 Grenoble, France}
\author{A.~Blanchet}
\affiliation{IRFU, CEA, Universit{\'e} Paris-Saclay, 91191 Gif-sur-Yvette, France}
\author{A.~Bonhomme}
\affiliation{IRFU, CEA, Universit{\'e} Paris-Saclay, 91191 Gif-sur-Yvette, France}
\author{C.~Buck}
\affiliation{Max-Planck-Institut f\"ur Kernphysik, Saupfercheckweg 1, 69117 Heidelberg, Germany }
\author{J.~Favier}
\affiliation{Univ. Grenoble Alpes, Universit{\'e} Savoie Mont Blanc, CNRS/IN2P3, LAPP, 74000 Annecy, France} 
\author{J.~Haser}
\affiliation{Max-Planck-Institut f\"ur Kernphysik, Saupfercheckweg 1, 69117 Heidelberg, Germany }
\author{V.~H\'{e}laine}
\affiliation{Univ. Grenoble Alpes, CNRS, Grenoble INP, LPSC-IN2P3, 38000 Grenoble, France}
\author{F.~Kandzia}
\affiliation{Institut Laue-Langevin, CS 20156, 38042 Grenoble Cedex 9, France }
\author{S.~Kox}
\affiliation{Univ. Grenoble Alpes, CNRS, Grenoble INP, LPSC-IN2P3, 38000 Grenoble, France}
\author{J.~Lamblin}
\email{jacob.lamblin@lpsc.in2p3.fr}
\affiliation{Univ. Grenoble Alpes, CNRS, Grenoble INP, LPSC-IN2P3, 38000 Grenoble, France}
\author{A.~Letourneau}
\affiliation{IRFU, CEA, Universit{\'e} Paris-Saclay, 91191 Gif-sur-Yvette, France}
\author{D.~Lhuillier}
\email{david.lhuillier@cea.fr}
\affiliation{IRFU, CEA, Universit{\'e} Paris-Saclay, 91191 Gif-sur-Yvette, France}
\author{ M.~Lindner} 
\affiliation{Max-Planck-Institut f\"ur Kernphysik, Saupfercheckweg 1, 69117 Heidelberg, Germany }
\author{L.~Manzanillas} 
\affiliation{Univ. Grenoble Alpes, Universit{\'e} Savoie Mont Blanc, CNRS/IN2P3, LAPP, 74000 Annecy, France}
\author{T.~Materna} 
\affiliation{IRFU, CEA, Universit{\'e} Paris-Saclay, 91191 Gif-sur-Yvette, France}
\author{A.~Minotti}
\affiliation{IRFU, CEA, Universit{\'e} Paris-Saclay, 91191 Gif-sur-Yvette, France}
\author{F.~Montanet} 
\affiliation{Univ. Grenoble Alpes, CNRS, Grenoble INP, LPSC-IN2P3, 38000 Grenoble, France}
\author{H.~Pessard}
\affiliation{Univ. Grenoble Alpes, Universit{\'e} Savoie Mont Blanc, CNRS/IN2P3, LAPP, 74000 Annecy, France} 
\author{J.-S.~Real} 
\affiliation{Univ. Grenoble Alpes, CNRS, Grenoble INP, LPSC-IN2P3, 38000 Grenoble, France}
\author{C.~Roca} 
\affiliation{Max-Planck-Institut f\"ur Kernphysik, Saupfercheckweg 1, 69117 Heidelberg, Germany }
\author{T.~Salagnac}  
\affiliation{Univ. Grenoble Alpes, CNRS, Grenoble INP, LPSC-IN2P3, 38000 Grenoble, France}
\author{S.~Schoppmann} 
\affiliation{Max-Planck-Institut f\"ur Kernphysik, Saupfercheckweg 1, 69117 Heidelberg, Germany }
\author{V.~Sergeyeva} 
\affiliation{Univ. Grenoble Alpes, Universit{\'e} Savoie Mont Blanc, CNRS/IN2P3, LAPP, 74000 Annecy, France}
\author{T.~Soldner}
\email{soldner@ill.fr}
\affiliation{Institut Laue-Langevin, CS 20156, 38042 Grenoble Cedex 9, France } 
\author{A.~Stutz} 
\affiliation{Univ. Grenoble Alpes, CNRS, Grenoble INP, LPSC-IN2P3, 38000 Grenoble, France}
\author{S.~Zsoldos}
\affiliation{Univ. Grenoble Alpes, CNRS, Grenoble INP, LPSC-IN2P3, 38000 Grenoble, France}

\collaboration{The \STEREO{} Collaboration}

% The "\note" macro will give a warning: "Ignoring empty anchor..."
% you can safely ignore it.

% e-mail addresses: only for the corresponding author


\date{\today}

\begin{abstract}
The reactor antineutrino anomaly might be explained by the oscillation of reactor antineutrinos toward a sterile neutrino of eV mass. In order to explore this hypothesis, the \STEREO{} experiment measures the antineutrino energy spectrum in six different detector cells covering baselines between 9 and 11 m from the compact core of the ILL research reactor. In this Letter, results from 66 days of reactor turned on and 138 days of reactor turned off are reported. A novel method to extract the antineutrino rates has been developed based on the distribution of the pulse shape discrimination parameter. The test of a new oscillation toward a sterile neutrino is performed by comparing ratios of cells, independent of absolute normalization and of the prediction of the reactor spectrum. The results are found to be compatible with the null oscillation hypothesis and the best fit of the reactor antineutrino anomaly is excluded at 97.5\% C.L.
\end{abstract}

%\keywords{}

\maketitle
%\linenumbers
Neutrino oscillation experiments of the last two decades have measured all mixing angles and mass splittings in a three flavor framework~\cite{GLOB}. Within this framework, no significant disappearance of neutrinos of few MeV energy is expected at baselines of less than 100~m. Nevertheless, many experiments at such baselines from nuclear reactors have observed a lower electron antineutrino flux than predicted. There are basically two possible explanations for this observation known as the reactor antineutrino anomaly (RAA)~\cite{RAA}. One is a deficient prediction of the antineutrino flux and spectrum from reactors, due to underestimated systematics of the measurements of beta spectra emitted after fission \cite{Feilitzsch1982,Schreckenbach1985,Hahn1989} or of the conversion method \cite{Mueller2011,Huber2011}, see \cite{Hayes2016,Huber2016} for recent reviews. The other one proposes new physics beyond the standard model of particle physics considering an oscillation from active toward a sterile neutrino state \cite{RAA}. The resulting disappearance probability for a neutrino of energy $E$ at distance $L$ from the source can be written as $\sin^2(2\theta_{ee}) \sin^2(\Delta m^2_{41} L/4 E)$ where $\theta_{ee}$ is the mixing angle and $\Delta m^2_{41}$ the difference of the mass squares of the mass eigenstates. This sterile neutrino option could also explain the deficits observed by the solar neutrino experiments GALLEX and SAGE in their calibrations with intense $^{51}$Cr and $^{37}$Ar neutrino sources \cite{Kaether2010,Abdurashitov2009,Giunti2011}. The original contours of allowed regions given in \cite{RAA} and their best fit values [$\sin^2 (2 \theta_{ee})=0.14$, $\Delta m^2_{41}$=2.4~eV$^2$] are used as a benchmark in this Letter. A recent review of light sterile neutrinos in this context and fits in different scenarios can be found in \cite{Giu}. In contrast, other experimental results strongly constrain oscillations to sterile neutrinos in different channels, putting tension on global fits~\cite{Dent}. In particular appearance and disappearance data appear incompatible.

Both explanations of the RAA can be studied with the data of the \STEREO{} experiment. \STEREO{} is installed at the High Flux Reactor of the Institut Laue-Langevin whose compact core (80~cm high, 40~cm diameter) operates with highly enriched $^{235}$U (93\%). Therefore, contributions from fission of other isotopes are negligible and \STEREO{} will provide a pure $^{235}$U antineutrino spectrum measured at a 10~m baseline. However, in this Letter we concentrate on the sterile neutrino hypothesis which has triggered a series of reactor antineutrino experiments at very short baselines \cite{Buck}. Results of the first two experiments, DANSS~\cite{DAN} and NEOS~\cite{NEOS}, exclude significant parts of the allowed region from Ref.~\cite{RAA}, but a combined analysis of all reactor antineutrino disappearance experiments still favors oscillations involving a fourth neutrino state at the 3$\sigma$ level~\cite{Dent}. The best fit parameters driven by the new DANSS and NEOS results suggest a mass splitting of $\Delta m^2_{14}\approx 1.3$~eV$^2$ and a mixing angle of $\sin^2 (2\theta_{ee})\approx0.05$, which is slightly outside the favored regions of Ref.~\cite{RAA} toward a lower mixing angle. This result is based on the comparison of purely spectral information. The analysis of DANSS compares the antineutrino spectrum of the movable detector for two baselines. However, it awaits calculation of the final systematic uncertainties~\cite{DAN}. NEOS relies on a nontrivial comparison of their data to the measured Daya Bay spectrum \cite{DB} obtained at different reactors with different detectors where the correction of the spectra requires inputs from predictions. Recently PROSPECT~\cite{PROSPECT} and NEUTRINO-4~\cite{NEUTRINO4} have presented first results.

In \STEREO{}, the antineutrino spectrum with energies up to about 10~MeV is measured in a segmented detector using six identical target cells of 37~cm length, whose centers are placed from 9.4 to 11.1~m from the reactor core. The sterile neutrino hypothesis can be tested by comparing the measured antineutrino energy spectra of the different cells. A neutrino oscillation with a mass splitting in the electron Volt region would manifest in a clear spectral pattern of a distance-dependent distortion of the energy spectrum. The analysis presented here uses spectra ratios with one cell as reference. It does not require a reactor spectrum prediction and is independent from the absolute flux normalization, minimizing systematic uncertainties.

\begin{figure}[h]
\centering
\includegraphics[width=0.48\textwidth]{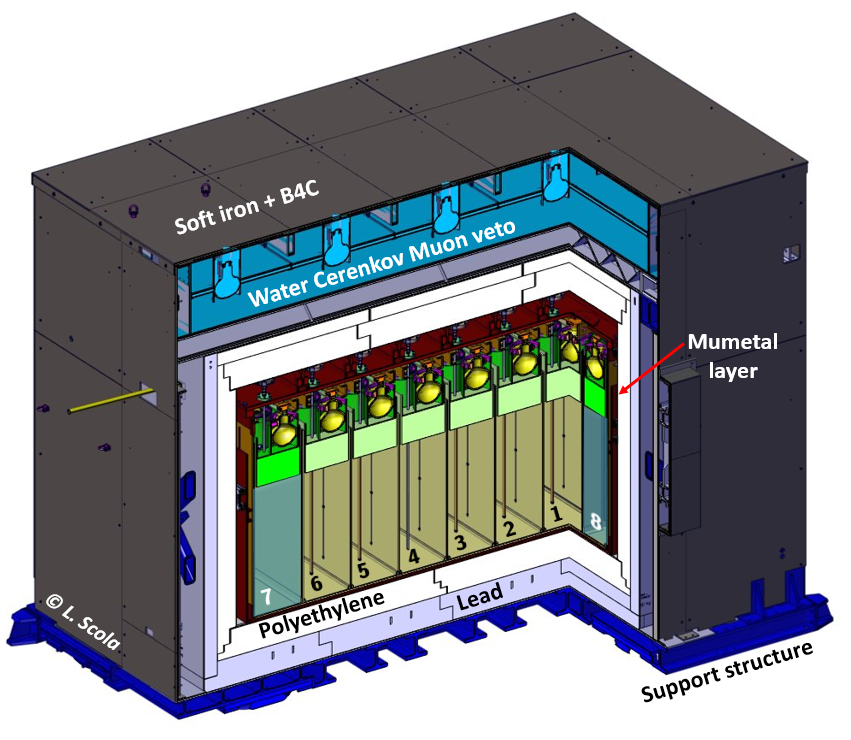}
\caption{\STEREO{} setup. 1--6: target cells (baselines from core: 9.4--11.1~m); 7, 8: two of the four gamma catcher cells.}
\label{fig:Detector}
\end{figure}

The \STEREO{} detector system~\cite{Ste} (see Fig.~\ref{fig:Detector}) consists of an antineutrino detector, a muon veto on top and several calibration devices. The antineutrinos are detected via the inverse beta decay reaction (IBD) on hydrogen nuclei in an organic liquid scintillator: $\bar{\nu}_{e}+p \rightarrow e^+ +n$. The six optically separated cells of the target volume are filled with a gadolinium (Gd) loaded liquid scintillator for a total of almost 2~m$^3$. They are read out from the top by four photomultiplier tubes (PMT) per cell. The IBD signature is a coincidence of a prompt positron and a delayed neutron capture event. The antineutrino energy is directly inferred from the prompt event as $E=E_{\text{prompt}}+0.782$~MeV. The neutron from the IBD reaction is moderated and then mainly captured by Gd isotopes. This capture creates a gamma cascade with about 8~MeV total energy. These gammas can interact in the target and in the gamma catcher, a segmented volume surrounding the target, filled with liquid scintillator without Gd and equipped with 24 PMTs. In some cases, the gamma catcher serves also for the total positron energy, detecting annihilation gammas escaping the target. The mean capture time of the coincidence signal is about 16~$\mu$s allowing for efficient discrimination of accidental background. Moreover, background events are strongly reduced by a thorough passive shielding design of various materials with a total mass of about 65~tons. \STEREO{} is installed underneath a water channel providing, together with the reactor building, an overburden of 15~m water equivalent against cosmic radiation. The remaining background can be measured during phases with the reactor turned off. A \textsc{Geant4}~\cite{Geant} (version 10.1) Monte Carlo model (MC) based on DCGLG4sim~\cite{DoubleChooz2012} describes detector geometry, shielding, position to the reactor core, particle interactions including neutron moderation and capture; light production, transport including cross talks between cells and detection, and signal conversion in the electronics. A method has been developed to convert the measured (or simulated) PMT signals into a reconstructed energy, taking into account light cross-talk between cells which ranges up to 15~\%. The reconstructed energy resolution ($\sigma/E$) for $^{54}$Mn $\gamma$ rays (0.835\,MeV) is about 9\%. Energy nonlinearity, due to quenching effects, is measured precisely and reproduced by the MC at the percent level. Drifts of the reconstructed energy are at the subpercent level. More information on the detector and its performances can be found in \cite{Ste}.
The analysis presented in this Letter concerns phase I of the experiment with 66~days of reactor turned on and 138~days of reactor turned off \cite{Data}.

Table \ref{tab:cuts} lists the set of IBD selection cuts corresponding to the best compromise between detection efficiency and background rejection, where the results remain quite stable around the chosen values. Beyond the basic cuts on energy and capture time (cuts 1-3), the detector segmentation is exploited to tag the topology of energy deposition of IBD events: compact prompt event only allowing for escaping 511~keV annihilation $\gamma$ rays (cuts 4 and 5), lower energy deposition threshold in the target for the expanded deposition pattern of the \textit{n}-Gd capture (cut 6), and upper distance threshold between the reconstructed vertices of prompt and delayed signals (cut 7). A 100\,$\mu$s muon veto (cut 8) and an isolation cut against multineutron cascades (cut 9) reject a large part of the cosmic-ray induced background. Untagged muons stopping and decaying in the top layer of the detector, without depositing more than 7.125~MeV energy, may be mistaken as IBD candidates. They are removed by the asymmetry of their light distribution between the PMTs of the vertex cell (ratio of maximum charge in a single PMT to total charge), which is larger than for events in the detector bulk (cut 10). The effects of these cuts on spectra and cell efficiencies are well described by the MC which was studied using measurements with sources as well as antineutrino runs. For example, measurements with an AmBe neutron source at various positions in the detector demonstrated that cell-to-cell differences in the data-to-MC ratio of the cut efficiencies were less than 1\%. These differences are included in the systematic uncertainties. The main contributions to the dead time are from the muon veto and isolation cuts. The total correction ranges from 10 to 15\% depending on the single rates induced by the activities of the neighboring experiments. It is accurately computed using two independent methods and leads to a relative uncertainty of 0.3\% over the data taking time.

\begin{table}[h]
	\centering
	\begin{tabular}{| l | c l| }
		\hline
		& & Applied cut\\
		\hline
		\hline
		Energy & (1) &  1.625\,MeV$< E_{\text{prompt}}<$ 7.125\,MeV \\
		& (2) & 4.5\,MeV$<E_{\text{delayed}}<$ 10\,MeV \\ 
		\hline
		Time & (3) &  0.25\,$\mu$s$<$ $\Delta \text{T}_{\text{prompt-delayed}}$ $<$ 70\,$\mu$s \\
		\hline
		Topology & (4) & $E_{\text{gamma catcher, prompt}} <$ 1.1\,MeV\\
		& (5) & $\forall~i~\neq~i_{\text{vertex}}$, $E_{i,\text{prompt}} <$ 0.8\,MeV\\
		& (6) &  $E_{\text{target, delayed}}>$ 1\,MeV\\
		& (7) &  $D_{\text{prompt-delayed}} <$ 600\,mm \\
		\hline
		Rejection of & (8) &  $>$ 100\,$\mu$s after a muon tag\\		
		$\mu$ induced & (9) &  No other event with $E>$ 1.5\,MeV \\
		background & & in $\pm$100\,$\mu$s window \\
		 %& (10) & $Q_{\text{max}}$/$Q_{\text{cell, prompt}} > $ 0.5\\
		 & (10) & $Q_{\text{PMT max, prompt}}$/$Q_{\text{cell, prompt}} \le $ 0.5\\
		\hline
	\end{tabular}
	\caption{Selection cuts for IBD-pair candidates.\label{tab:cuts}}
\end{table}

\begin{figure}[h]
\centering
\includegraphics[width=0.50\textwidth]{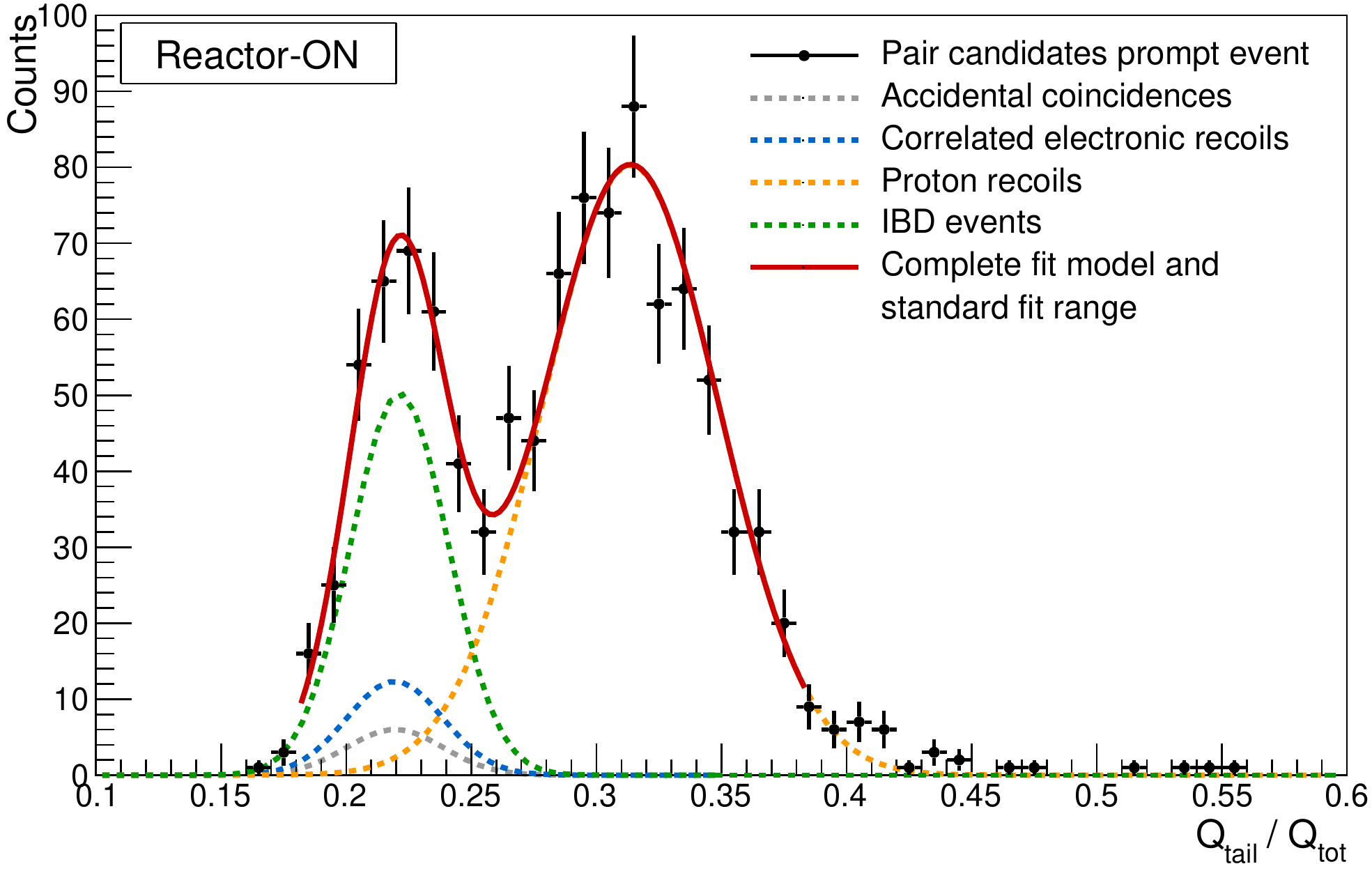}
\caption{Example of the PSD distribution for events in cell 1 with reconstructed energy in [3.125,3.625] MeV, collected in 22.8 days of reactor turned on. The dashed curves illustrate the four components of the model.}
\label{fig:PSD_Components}
\end{figure}

The antineutrino signal is separated from the remaining background using a pulse shape discrimination (PSD) parameter defined as the ratio of the pulse tail to total charge. The PSD distribution of the prompt event of all pair candidates passing the IBD selection cuts is shown in Fig.~\ref{fig:PSD_Components} for one of the eleven reconstructed energy bins defined in the analysis. Two classes of events clearly appear, the proton recoils due to muon-induced fast neutrons at high PSD and the electronic recoils at low PSD. The electronic recoil class comprises IBD events, correlated electronic background induced by cosmic rays, and accidental coincidences (the single rates being dominated by gammas).

After splitting the data into time bins of 1 week and energy bins of 500~keV width for each cell, the PSD distribution of each bin is modeled as sum of a proton recoil, an electron recoil and -- for reactor-on data -- an IBD Gaussian. Area $A_{\rm p}$, position and width of the proton recoil Gaussian are determined directly from the fit. Position and width of the electronic recoil Gaussian, comprising accidentals and correlated events, are fixed to the values $\mu_{\rm s}$ and $\sigma_{\rm s}$ of the PSD distribution of singles obtained with negligible statistical uncertainty. The area can be separated into $A_{\rm el}=A_{\rm acc}+A_{\rm p} R_{\rm cosmic}$. $A_{\rm acc}$ is determined by a simultaneous fit to the PSD distribution of accidental events, extracted with high statistics by looking for random delayed events in many (typically 100) delayed windows for each prompt candidate and rescaling by the number of windows. This accounts correctly for changing uncorrelated background. $R_{\rm cosmic}$ parametrizes the ratio of correlated electronic recoils to proton recoils. Whereas the rates of both event types depend on atmospheric pressure, their ratio was found to be compatible with a constant. This can be understood since, within the applied cuts, electronic recoils in the prompt event are dominantly created by primary \mbox{(multi)neutron} spallation, e.g., via $^{\rm 12}{\rm C}(n,n')\gamma$ or gammas of double neutron capture events. $R_{\rm cosmic}$ is determined from reactor-off data for each energy bin and its time average and statistical uncertainty are the only parameters transferred to the analysis of the reactor-on data, as a pull term in the PSD fits. The PSD distribution of IBD prompt events is slightly different from that of singles because of the positron annihilation gammas. This is accounted for by constraining position and width of the IBD Gaussian only moderately in the fit, to $\mu_{\rm s}\pm 0.2\sigma_{\rm s}$ and $(0.95\pm0.10)\sigma_{\rm s}$, respectively. These constraints have been estimated from the difference of the reactor-on and reactor-off PSD distributions. Finally, the area of the IBD Gaussian $A_{\rm IBD}$ yields the number of antineutrinos for the respective time-energy bin.

In contrast to a fixed cut on the PSD value, this novel method permits a full separation of the different contributions to electronic and proton recoils, in spite of the overlapping distributions, and accounts for slow drifts in the PSD distribution. The method is insensitive to dead time differences between reactor-on and reactor-off runs since rates entering in the ratios are measured simultaneously and only ratios are transferred between reactor-on and reactor-off measurements. The remaining systematics due to deviations of the model from the true PSD distribution is controlled by the high goodness of fit and the stability with respect to the fit ranges for all energy bins of reactor-off PSD distributions. Moreover, since this model is applied to all cells, potential deviations from the model will be further suppressed in the ratio of spectra used in the oscillation search.

The resulting total antineutrino rate is $(396.3\pm 4.7)\,\bar{\nu}_e$/day with a signal to background ratio of about 0.9, determined from integrating over the region of interest in the PSD parameter. To search for a possible oscillation toward a sterile neutrino in the data, a ratio method is used. It consists of dividing bin by bin the spectrum of cells 2 to 6 by the spectrum of cell 1, which serves as reference, and comparing these ratios between data and MC. This formalism is insensitive to the model of the reactor spectrum and relies only on the relative difference between cells. However, the variance of the ratio is not well defined when the denominator approaches zero within few $\sigma$ units. Therefore, this analysis has been limited to $E_{\text{prompt}}<$7.125\,MeV. In this range, the smallest denominator value is 4.7 $\sigma$ away from zero. A profile $\Delta \chi ^2$ method is used with
\begin{widetext}
\begin{equation}\label{eqn:chi2}
\chi ^2 = \sum_{i=1}^{N_{\text{ebin}}}\bigg(\overrightarrow{R_i^{\text{data}}} - \overrightarrow{R_i^{\text{MC}}(\alpha)} \bigg)^t V_i^{-1}\bigg(\overrightarrow{R_i^{\text{data}}} - \overrightarrow{R_i^{\text{MC}}(\alpha)} \bigg)  + \sum_{l=1}^{N_{\text{cells}}} \left( \frac{\alpha_l^{\text{norm}}}{\sigma_l^{\text{norm}}}\right)^2  +  \sum_{l=0}^{N_{\text{cells}}} \left( \frac{\alpha_l^{\text{escale}}}{\sigma_l^{\text{escale}}}\right)^2 
\end{equation}
\end{widetext}

\noindent $\overrightarrow{R_i^{\text{data}}}$ and $\overrightarrow{R_i^{\text{MC}}(\alpha)}$ are five-dimensional vectors (cell 2 to cell 6) corresponding to the measured and the MC ratios, respectively, for the \textit{i}th energy bin. 
The MC takes into account the spatial distribution of IBD events for antineutrinos from the reactor core, the energy nonlinearities and the applied cuts in order to simulate the expected energy spectra. Since the energy spectrum of cell 1 is used as a common denominator for all ratios, the \textit{i}th energy bins of all ratios are highly correlated. This effect is coded in the covariance matrices $V_i$, whose off-diagonal elements have been determined by random sampling considering Gaussian uncertainties for the antineutrino rates of each bin. Nuisance parameters $\alpha$ are added to take into account systematic uncertainties: $\alpha_l^{\text{norm}}$ are the relative normalizations of the cells due to the uncertainties on the volume and detection efficiencies ($\sigma_{l \neq 4}^{\text{norm}}=1.7\%$ and $\sigma_4^{\text{norm}}=3.4\%$ because of reduced optical coupling for cell 4, see \cite{Ste}),  $\alpha_{l>0}^{\text{escale}}$ are the uncorrelated energy scale uncertainties driven by the cellwise residual discrepancies between the energy response of data and MC ($\sigma_{l>0}^{\text{escale}}=1.1\%$) and $\alpha_{0}^{\text{escale}}$ corresponds to the energy scale bias common to all cells due to the timewise evolution of the energy response ($\sigma_{0}^{\text{escale}}=0.35\%$). They enter into $R_{l,i}^{\text{MC}}(\alpha)$ as follows:
\begin{widetext}
\begin{equation}
R_{l,i}^{\text{MC}}(\alpha)= \frac{T_{l,i}}{T_{1,i}}\bigg( 1+\alpha_l^{\text{norm}}-\alpha_1^{\text{norm}}+\frac{\Delta T_{l,i} (\alpha_0^{\text{escale}},\alpha_l^{\text{escale}})}{T_{l,i}} -\frac{\Delta T_{1,i} (\alpha_0^{\text{escale}},\alpha_1^{\text{escale}})}{T_{1,i}} \bigg)
\end{equation}
\end{widetext}
where $T_{l,i}$ are the predicted spectra including oscillation and detector response and $\Delta T_{l,i}$ describe the changes obtained from neighbor energy bins depending on the energy scale parameters.

\begin{figure}[h]
\centering
\includegraphics[width=0.50\textwidth]{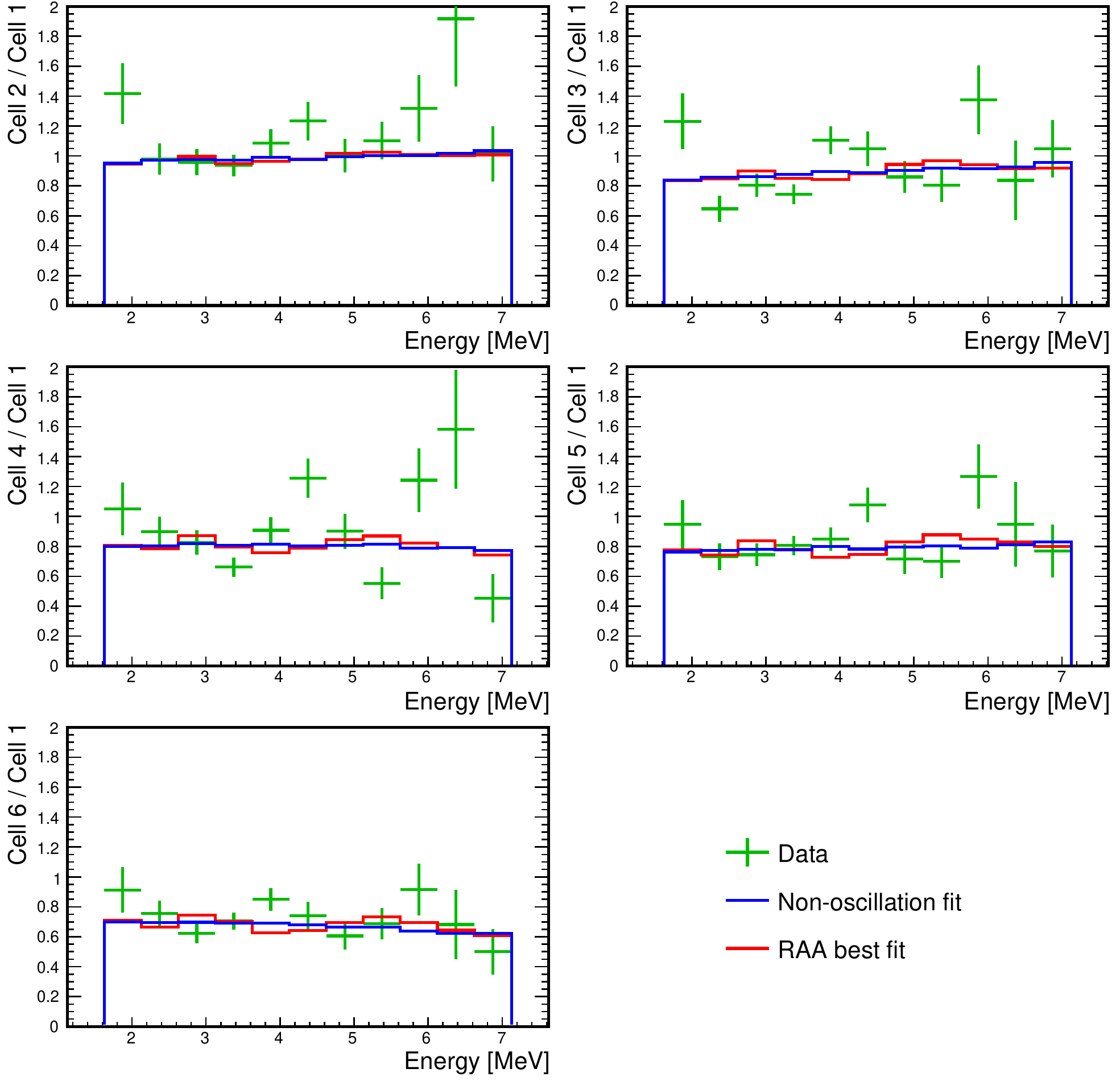}
\caption{Measured ratios cell $i$/cell 1 compared to the null oscillation hypothesis and the RAA best fit benchmark from \cite{RAA}. Energy is the reconstructed energy of the prompt event.}
\label{fig:Ratio}
\end{figure}

First, the null oscillation hypothesis has been tested. Fig.~\ref{fig:Ratio} compares the measured and the simulated ratios without oscillation (and with oscillations with the RAA best fit values from \cite{RAA}) after minimization with free nuisance parameters. The decrease of the mean value of the ratios with increasing distance reflects the $1/L^2$ flux dependence, where the cell detection efficiencies have to be taken into account. This dependence is confirmed quantitatively since the fitted cell normalization parameters $\alpha_l^{\rm norm}$ were found within the expected uncertainties. The simulated ratios are not perfectly flat because the energy response can slightly vary from one cell to another. From the probability density function of $\Delta \chi^2$ obtained by generating a large number of pseudoexperiments, the $\Delta \chi^2$ of 9.1 with respect to the minimum in the $(\sin^2 (2\theta_{ee}), \Delta m^2_{41})$ plane corresponds to a \textit{p} value of 0.34. Hence, the null oscillation hypothesis cannot be rejected.

\begin{figure}[h!]
\centering
\includegraphics[width=0.50\textwidth]{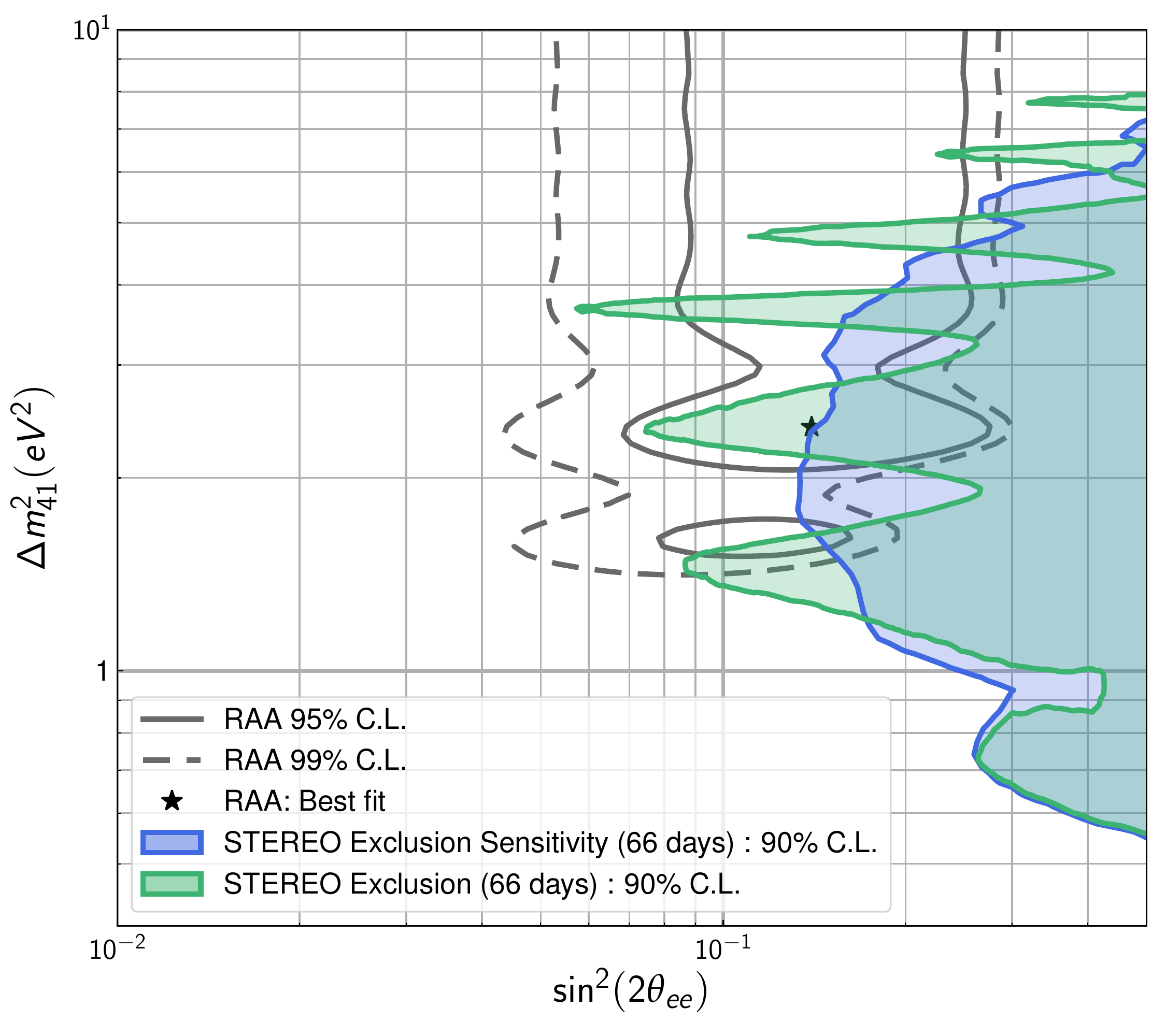}
\caption{Exclusion contour of the oscillation parameter space. The RAA values and contours are from \cite{RAA}.}
\label{fig:Contour}
\end{figure}

To infer an exclusion contour in the oscillation parameter space, a raster scan method \cite{Feldman} has been used. It consists of dividing the 2D parameter space into slices, with one slice per $\Delta m_{14}^2$ bin, and computing for each slice the $\chi^2$ as a function of $\sin^2 (2 \theta_{ee})$ with free nuisance parameters. Then, the $\Delta \chi^2$ values are computed using the minimum value of each slice and not the global minimum. The 90\% C.L. exclusion contour corresponds to the parameter space where the $\Delta \chi^2$ is higher than the value giving a one sided \textit{p} value of 0.1 in the probability density function obtained from pseudoexperiments for each bin of the parameter space.
The result is shown in Fig.~\ref{fig:Contour}. The exclusion contour is centered around the sensitivity contour, also computed with a raster scan, with oscillations due to the statistical fluctuations.
The original RAA best fit is excluded at 97.5\% C.L.

These first results demonstrate the ability of the \STEREO{} experiment to detect antineutrinos above the residual background, dominated by cosmic-ray induced events. With the novel method presented in this Letter, the proton recoil component of this background is measured in the temperature and pressure conditions of the reactor-on data taking while the associated relative contamination of electronic recoils is well constrained from the reactor-off data.
The accuracy of the background subtraction is thus driven by the statistics and improves with more reactor-off data acquired. The \STEREO{} data taking is in progress and should reach the envisaged statistics, 300 days at nominal reactor power, before the end of 2019.

We would like to thank G. Mention for discussions on the implementation of the ratio method. This work is supported by the French National Research Agency (ANR) within the project ANR-13-BS05-0007 and the programs P2IO LabEx (ANR-10-LABX-0038) and ENIGMASS LabEx (ANR-11-LABX-0012). We acknowledge the support of the CEA, CNRS/IN2P3, the ILL and the Max Planck Gesellschaft.
%\end{acknowledgments}

%\bibliography{FirstResultsSTEREO}

\end{document}